\documentclass{article}

\begin{document}

\title{Wavefunction Collapse and Conservation Laws}
\author{Philip Pearle\\
Hamilton College\\
Clinton, NY 13323}
\date{}
\maketitle

\begin{abstract}

	It is emphasized that the collapse postulate of standard quantum theory can violate 
conservation of energy-momentum and there is no 
indication from where the energy-momentum comes or to where it goes.   
Likewise, in the Continuous Spontaneous Localization  (CSL) dynamical collapse model, 
particles gain energy on average.  
In CSL, the usual Schr\"odinger dynamics 
is altered so that a randomly fluctuating classical field 
interacts with quantized particles to cause wavefunction collapse.  In this paper it is 
shown how to define energy for the classical field so that the average value 
of the energy of the field plus the quantum system {\it is}  
conserved for the ensemble of collapsing wavefunctions. While conservation of 
just the first moment of energy is, of course, 
much less than complete conservation of energy, this does support the idea that 
the field could provide the conservation law balance when events occur.  

\end{abstract}

\section{The Collapse Postulate and Nonconservation of Energy-Momentum}\label{Postulate}

\hspace{\parindent}It is not generally appreciated that the collapse postulate of standard quantum theory (SQT) 
can violate the geometric conservation laws, e.g., the conservation of energy and momentum\cite{PearleAcadSci}. 

	In quantum theory, conservation of a physical quantity represented by the operator $Q$ is the statement 
that the probability distribution of $Q$'s eigenvalues $q$, $|\langle q|\psi,t\rangle|^{2}$, is constant in time. 
For the Schr\"odinger evolution of a statevector, conservation is guaranteed by the vanishing of the commutator 
of $Q$ with the Hamiltonian.  This ensures that 
$\langle \psi,t|F(Q)|\psi,t\rangle=\langle \psi,0|F(Q)|\psi,0\rangle$, where $F$ is an arbitrary 
function of $Q$: the constancy of $|\langle q|\psi,t\rangle|^{2}$ immediately follows.  
(It will be useful for later purposes to note here that 
this is also equivalent to constancy of the expectation value of arbitrary integer powers of $Q$, 
which in turn is equivalent to constancy of the expectation value of the generating function $G\equiv \exp iaQ$.)

	However, according to SQT, the Schr\"odinger evolution is only part of a statevector's evolution.  
Under some circumstances (such as measurement) the ``collapse" postulate is to be applied: when
the states $|\psi_{i}\rangle$ become sufficiently ``macroscopically distinct," 
the statevector $|\psi,T\rangle=\sum_{i}\alpha_{i}|\psi_{i}\rangle$ is to be 
replaced by one of the orthonormal statevectors $|\psi_{i}\rangle$ with probability $|\alpha_{i}|^{2}$. 
Conservation of $Q$ after the collapse requires 
$|\langle q|\psi,T\rangle|^{2}=\sum_{i}|\alpha_{i}|^{2}|\langle q|\psi_{i}\rangle|^{2}$.  
Now, no one has ever been successful
 in precisely formulating the collapse postulate but, one 
might think, a careful formulation could be consistent with energy conservation.  In this section 
we give some examples where the collapse postulate is inconsistent with energy conservation, for all but a 
set of measure zero of superpositions of macroscopically distinct states. 
Actually, conservation of energy and momentum in one reference frame are both needed to have conservation 
of energy in all frames. Thus we shall look for both energy and momentum conservation 
to be satisfied by the collapse.  As we shall see, momentum conservation turns out to be 
a more stringent condition than energy conservation.  

	For a simple example, consider a 
one-dimensional macroscopic ``pointer" with statevector $|\Psi\rangle=|\Phi\rangle|\phi\rangle$, where 
$|\Phi\rangle$ describes its center of mass (cm) and $|\phi\rangle$ describes its internal degrees of freedom.
If $|\Phi\rangle=\alpha_{1}|\Phi_{1}\rangle +\alpha_{2}|\Phi_{2}\rangle$, where the $\alpha_{i}$ are arbitrary 
but the orthogonal wavefunctions $\langle x |\Phi_{1}\rangle$, $\langle x |\Phi_{2}\rangle$ are "sufficiently" 
distinctly localized (e.g., gaussian-like peaks which are narrow and well-separated, 
but with small wiggles in the right places in the tails so as to to be orthogonal), then the collapse postulate says that 
$|\Psi\rangle$ is to be replaced by one of the $|\Phi_{i}\rangle|\phi\rangle$ with probability $|\alpha_{i}|^{2}$. 
However, we shall now point out that the class of distinctly localized pairs of states 
\{$|\Phi_{1}\rangle,|\Phi_{2}\rangle \}$ 
for which energy-momentum is conserved in collapse are a set of measure zero.  

	  Conservation of energy-momentum requires 
the expectation value of the generating functions for energy and momentum to be unchanged by the collapse.
These generating functions are, respectively, $G_{E}\equiv\exp ia(H_{cm}+H_{int})\equiv G_{cm}G_{int}$  
($H_{cm}\equiv P^{2}/2M$ where $P$ is the cm momentum operator and $M$ is the pointer mass; $H_{int}$ is the 
Hamiltonian of the internal degrees of freedom) and $G_{P}\equiv\exp ibP$, 
 
	Then, $\langle\Psi |G_{P}|\Psi\rangle=\sum_{i=1}^{2}|\alpha_{i}|^{2}\langle\Phi_{i}|G_{P}|\Phi_{i}\rangle$ 
is implied by momentum conservation in collapse,  
i.e., $\alpha ^{*}_{1}\alpha_{2}\langle\Phi_{1}|G_{P}|\Phi_{2}\rangle+
\alpha_{1}\alpha^{*}_{2}\langle\Phi_{2}|G_{P}|\Phi_{1}\rangle =0$.  Since the collapse 
postulate is to be applied to the states $|\Phi_{i}\rangle$ because of their macroscopic distinctness 
and regardless of the amplitudes $\alpha_{i}$, the phase of 
$\alpha_{1}\alpha ^{*}_{2}$ is arbitrary, so $\langle\Phi_{2}|\exp ibP|\Phi_{1}\rangle=0$.  
Multiplying this by $\exp -ibp$ and integration over b gives the momentum projection operator $\delta(P-p)$, 
with the result 

\begin{equation}\label{1}
 \langle \Phi_{2}|p\rangle\langle p|\Phi_{1}\rangle =0
\end{equation} 

	The similar argument for energy implies that 
$\langle\Phi_{2}|G_{cm}|\Phi_{1}\rangle \langle\phi |G_{int}|\phi \rangle =0$.  However, 
if Eq. (\ref{1}) is satisfied then the first factor vanishes: 

\begin{eqnarray*}
\langle\Phi_{2}|G_{cm}|\Phi_{1}\rangle=\int dp\langle\Phi_{2}|p\rangle\langle p|\Phi_{1}\rangle e^{iaP^{2}/2M}=0.     
\end{eqnarray*}

\noindent Therefore, (\ref{1}) implies not only momentum conservation but 
also energy conservation in collapse as well. 
 
	The condition (\ref{1}) is very stringent: most (all but a set of measure zero) 
superpositions of macroscopically distinct states will not satisfy it. But, actually, 
of those states which {\it do} satisfy (\ref{1}) only a measure zero set of {\it them} will 
be macroscopically distinct. For, in order that (\ref{1}) be satisfied, 
there must be at least one range of momentum 
$p_{1}\leq p\leq p_{2}$ over which $\langle p|\Phi_{2}\rangle=0$ and

\begin{eqnarray*}
\langle p|\Phi_{1}\rangle = \Theta (p_{2}-p)\Theta (p-p_{1})\langle p|\Phi\rangle    
\end{eqnarray*}

\noindent ($\Theta$ is the step function).  However, each such range of momentum contributes a piece 
to the wavefunction $\langle x|\Phi_{1}\rangle$ that has infinite $\langle x^{2}\rangle$: 

\begin{eqnarray*}
\langle x^{2}\rangle \!\!\!\!\!\!\!\!\!\!&&=\int_{-\infty}^ {\infty}dp\langle\Phi_{1}|p\rangle 
(i\frac{d}{dp})^{2}\langle p|\Phi_{1}\rangle =
\int_{-\infty}^ {\infty}dp|\frac {d}{dp}\langle p|\Phi_{1}\rangle|^{2}\\
&&=|\langle p_{1}|\Phi\rangle|^{2}\int_{-\infty}^ {\infty}dp\delta^{2}(p-p_{1})+
|\langle p_{2}|\Phi\rangle|^{2}\int_{-\infty}^ {\infty}dp\delta^{2}(p-p_{2})+...=\infty
\end{eqnarray*}
  
	Another way to see this is to note that such a momentum space wavefunction segment 
contributes an asymptotic $x^{-1}$ behavior to the position space wavefunction $\langle x|\Phi_{1}\rangle$,  
since integrating the identity

\begin{eqnarray*}
e^{ipx}\Phi (p)=\frac{d}{dp}\{e^{ipx}[\frac{\Phi (p)}{ix}-\frac{\Phi '(p)}{(ix)^{2}}+
\frac{\Phi ''(p)}{(ix)^{3}}-...]\}
\end{eqnarray*} 

\noindent results in 

\begin{eqnarray}\label{2}
\int_{p_{1}}^{p_{2}}dpe^{ipx}\langle p|\Phi\rangle=
e^{ip_{2}x}\frac{\langle p_{2}|\Phi\rangle}{ix}[1+\hbox{O}\frac{1}{x}]-
e^{ip_{1}x}\frac{\langle p_{1}|\Phi\rangle}{ix}[1+\hbox{O}\frac{1}{x}]
\end{eqnarray}

	 This demonstration that the wavefunctions are not well localized 
assumes that $|\langle p_{i}|\Phi\rangle|^{2}\neq 0$. A set of measure zero 
of {\it these} wavefunctions 
has zeros at the $p_{i}$'s.  Then, a similar argument shows that $\langle x^{4}\rangle$ 
is infinite provided that also $\langle p_{i}|\Phi\rangle '\neq 0$, etc. We see 
therefore that one {\it can} construct 
a superposition of distinctly localized wavefunctions which conserve energy-momentum 
in collapse if the zeros at the $p_{i}$'s are of large enough order. An example is if the momentum space form is 

\begin{eqnarray*}
\langle p|\Phi_{1}\rangle\sim\Theta (p_{2}-p)\Theta (p-p_{1})(p_{2}-p)^{n}(p-p_{1})^{n}\exp -ipb 
\end{eqnarray*}

\noindent (where $n$ is large) and similarly for $\langle p|\Phi_{2}\rangle$ (with nonoverlapping $p_{i}$ 
ranges, and a quite different mean position $b$).

	We conclude that all but a set of measure zero of 
wavefunctions $\langle x|\Phi_{i }\rangle$ which are macroscopically distinct 
do not conserve energy-momentum under the collapse rule. 

	To push this a bit further, given the desideratum of energy-momentum conservation, one could imagine 
a collapse rule which is not of the usual type, in that it would not require collapse of 
$|\Phi\rangle=\alpha_{1}|\Phi_{1}\rangle +\alpha_{2}|\Phi_{2}\rangle$ for {\it arbitrary} $\alpha_{i}$, but
 only for one set of $\alpha_{i}$, provided energy-momentum is conserved.  Then one cannot invoke the 
 arbitrariness of phases used to obtain (\ref{1}).  Instead, one may without loss of generality absorb phases 
 in the $|\Phi_{i}\rangle$ so that $\alpha_{1}$ is real and $\alpha_{2}$ is imaginary.  Then, (\ref{1}) is 
 replaced by the more general energy-momentum conserving requirement
 
\begin{eqnarray*}
 \langle \Phi_{2}|p\rangle\langle p|\Phi_{1}\rangle =\langle \Phi_{1}|p\rangle\langle p|\Phi_{2}\rangle 
\end{eqnarray*}

\noindent This can be satisfied by wavefunctions which obey (\ref{1}) with the consequences presented above  
but, in addition, it can be satisfied if the momentum space wavefunctions have identical phases: 

 \begin{eqnarray*}
\langle p|\Phi_{i}\rangle = R_{i}(p)e^{i\theta (p)}
\end{eqnarray*}

\noindent ($R_{i}$, $\theta$  are real).  Certainly this is a set of measure zero but,  
more than that, it is hard to see how such wavefunctions could be of the macroscopically distinct type 
one desires.  To see this, consider that the criterion $\langle \Phi_{2}|\exp iPb|\Phi_{1}\rangle=
\langle \Phi_{1}|\exp iPb|\Phi_{2}\rangle$ is, in position space, 

\begin{eqnarray*}
\int dx\Phi_{2}^{*}(x) \Phi_{1}(x+b) = \int dx\Phi_{2}(x) \Phi_{1}^{*}(x-b)
\end{eqnarray*}

\noindent Thus the absolute magnitude of the overlap integral of one wavefunction with 
the other translated by the distance $b$ is the same for translation to the right or left.  
Since the integral = 0 for $b=0$ (the wavefunctions are orthogonal) 
and vanishes for infinite $b$, it must have a maximum 
for some translation to the right and for an equal translation to the left.  It is hard to reconcile that 
behavior with distinctly localized wavefunctions whose overlap integral one would think would grow to 
a maximum for translation in one direction as the peaks of the wavefunctions are brought to overlap, 
but which would diminish for translation in the other direction 
as the peaks are moved further away from each other. 

	So, we again conclude that all but a set of measure zero of 
wavefunctions which are macroscopically distinct do not conserve energy-momentum, 
even under this relaxed collapse rule. 

	Of course, the above example is rather specialized.  A practical 
scheme for getting a pointer into a superposition of two spatially separated states could very well entail 
different internal states for the two pointer positions and entail correlation with a system 
external to the pointer as in a measurement.  However, that a similar difficulty exists 
in realistic cases can be seen as follows\cite{Goldberg}.

	Consider the measurement of whether a particle is in a given region (again, for simplicity we will discuss 
just a one dimensional world: the argument is easily extended to three dimensions) so that the apparatus 
will either register ``in the region" (1) or ``outside the region" (2).  The apparatus may be as 
complicated as needed, it may not make a perfect measurement, it may disturb the particle: 
we just suppose that the final state of apparatus and particle at time $T$ is a superposition of two orthogonal 
macroscopically distinguishable states
$|\Psi,T\rangle = \alpha_{1}|\chi_{1}\rangle+\alpha_{2}|\chi_{2}\rangle$. We also suppose, going backwards 
in time, that $\exp iH(T-T_{0})|\chi_{i}\rangle=|\Phi_{i}\rangle|\Gamma\rangle|\gamma\rangle$, where 
$|\Phi_{i}\rangle$ is a state of the particle at initial time $T_{0}$ and $|\Gamma\rangle$, $|\gamma\rangle$ are 
respectively the initial apparatus cm and internal statevectors, so 
 
\begin{eqnarray*}
|\Psi, T_{0}\rangle=[\alpha_{1}|\Phi_{1}\rangle+\alpha_{2}|\Phi_{2}\rangle]|\Gamma\rangle|\gamma\rangle
\end{eqnarray*}

	Now, suppose we adopt the collapse postulate to 
collapse the final statevector to states $|\chi_{i}\rangle$ and that 
energy-momentum is conserved. In particular, this means that 
$\langle \chi_{2}|\exp ib(P+P_{cm})|\chi_{1}\rangle=0$, where $P_{cm}$ is the apparatus 
momentum operator (which only acts on the apparatus cm statevector) and $P$ is the particle's 
momentum operator.  Since $[P+P_{cm},H]=0$, this means that 

\begin{eqnarray*}
\langle\Phi_{2}| e^{ibP}|\Phi_{1}\rangle\langle\Gamma| e^{ibP_{cm}}|\Gamma\rangle
\langle\gamma|\gamma\rangle=0
\end{eqnarray*} 

\noindent For a reasonable initial apparatus cm wavefunction (such as a gaussian), the second scalar product  
(of $|\Gamma\rangle$ with itself translated by a distance $b$) does not vanish for any finite $b$.  This 
means that $\langle\Phi_{2}| e^{ibP}|\Phi_{1}\rangle = 0$, which implies that (\ref{1}) and therefore 
(\ref{2}) are satisfied by the initial {\it particle} wavefunction. 

	Thus, the hypothesis that energy-momentum is conserved in 
the collapse to the states $|\chi_{i}\rangle$ requires (apart from the caveat about a set of measure zero 
discussed previously) that the apparatus 
measures one infinite $\langle x^{2}\rangle$ state of the particle as being inside the 
region and another infinite $\langle x^{2}\rangle$ state as being outside the region. This could not 
be the case for a reasonably designed (even somewhat imperfect) apparatus. 
One must conclude that the hypothesis is untenable and 
that energy-momentum will generally not be conserved if the collapse postulate is applied.

\section{A Collapse Model and Nonconservation\\ of Energy-Momentum}\label{Intro}

\hspace{\parindent}Dynamical collapse models replace the collapse postulate of SQT by a modified 
Schr\"odinger equation which describes wavefunction collapse as a 
continuous physical process. The hope is that there really is such a process  
and that construction of phenomenological models and 
investigation of their experimental consequences will contribute 
to its confirmation. The first such models \cite{BohmBub,Pearle76-86} were 
designed to produce, as final states, just the results of the collapse postulate.  
Thus their conservation law violation is just of the type illustrated above.  
However, these models have the unsatisfactory feature that the onset of collapse 
and the ``preferred basis" (final states of collapse) are put in 
by hand, for each application\cite{PearlePerugia, Pearle89}. 
The Spontaneous Localization 
(SL) model\cite{GRW,GRReview} of Ghirardi, Rimini and Weber 
showed how to overcome this, although its 
method of achieving collapse is not via a modified Schr\"odinger equation, 
and it has the unsatisfactory feature of violating particle 
exchange symmetry. These last problems are overcome, and the good features of 
earlier models and SL are retained in the nonrelativistic 
Continuous Spontaneous Localization (CSL) 
model\cite{Pearle89,GPR,GPReview,PearleNaples}.  However, SL and CSL introduce 
a new mechanism for conservation law violation.  

	In CSL a randomly fluctuating classical field $w({\bf x},t)$ interacts with the 
particle number density (or mass density or energy density) operator 
to produce collapse toward its spatially localized eigenstates (this resolves the 
preferred basis problem).  CSL possesses the SL feature that the collapse 
interaction is always "on" (this resolves the collapse onset problem). The collapse 
of a many--particle state in a superposition of widely separated clumps to one of 
the clumps is rapid, but even a single isolated particle continually undergoes collapse, 
a narrowing of its wavefunction, albeit slowly.  This narrowing means that its 
energy increases.  

	The predicted increase of particle energy due to collapse has been the 
focus of experimental tests\cite{PearleSquires1,Collett,Ring} (in lieu 
of the more difficult direct tests of macroscopic 
interference\cite{PearleZ,Leggett,Clauser}).  These have suggested that 
the coupling between $w({\bf x},t)$ and the particle number density 
operator is proportional to the particle's mass, i.e., that $w({\bf x},t)$'s coupling is to mass density 
(or energy density), with its suggestive overtones of a connection 
between collapse and gravity\cite{Karolyhazy,Penrose,Diosi,GGR,PearleSquires2}.

	This violation of the conservation of particle energy has also been the
focus of criticism of the CSL model\cite{Ballentine, Anandan}. 
As has been emphasized in section 1, this criticism also deserves to 
be applied to SQT plus the collapse postulate.  However, here 
I shall show a way to define energy for the complete system of classical field plus quantized particles 
so that its expectation value is constant, not for 
any individual statevector $|\psi,t\rangle_w$ 
evolving under its particular field $w({\bf x},t)$, but  
for the ensemble of collapsing wavefunctions $\{ |\psi,t\rangle_w\}$ (with the correct 
probability for the occurrence of each $w$, as given by CSL).  This 
is as it should be: the ensemble of collapsed states and their associated fields  
$w$ describe the realized physical states in nature, 
and it is this ensemble which should satify the conservation law.
   
	Of course this is far from complete conservation of energy: that 
would require conservation of all powers of the energy, not just the first.  Perhaps 
conservation of the energy expectation value is as much as one might expect 
from a model where the collapse--causing field isn't quantized.  However, 
this result does suggest that the hitherto unaccounted for 
violation of energy conservation, by SQT plus the collapse postulate,  
involved in describing the occurrence of physical events,  
may be accounted for in a dynamical collapse model as due to 
an energy exchange with a collapse--causing field.
  
  It should be mentioned that a number of authors have suggested models 
where collapse takes place toward energy eigenstates (rather than toward energy 
density eigenstates as in CSL)\cite{SouthAfricans, Percival, Hughston}. In such models  
energy is conserved but the resulting collapsed states 
may not be the states seen in nature: macroscopic superpositions of 
spatially separated states of the same energy result from the dynamics of these models.  This 
seems to miss the point of a collapse model which is, roughly speaking, 
``what you see (in nature) is what you get (from the theory)".   

	For explanatory ease,  
a simpler CSL model than 
described above shall be employed.  In this model 
the collapse is toward the eigenstates 
of a single operator $A$ rather than nonrelativistic CSL's 
collapse toward the eigenstates of 
the mass density or energy operator (actually an infinity of commuting operators, 
one at each point of space). This uses a fluctuating classical quantity $w(t)$ which 
only depends upon $t$.  Section 3 contains a review of this CSL formalism.  
Section 4 presents the expression for the energy associated with $w(t)$ 
plus the quantum system, 
and gives the proof of conservation of the ensemble mean energy.  
In section 5, expressions are given separately for 
the quantum system's mean energy and $w(t)$'s mean 
energy, showing how a change of the former is at the expense of the latter. In conclusion, 
section 6 contains a simple example and 
also sketches how to apply this to the full 
nonrelativistic CSL model and to the other geometric conservation laws such as 
momentum conservation.

\section{CSL}\label{CSL}

\hspace{\parindent}Consider the statevector evolution  
	
\begin{equation}\label{3}
	|\psi,T\rangle_{w}\equiv{\cal T}e^{-\frac{1}{4\lambda}\int_{0}^{T}
 dt[w(t)-2\lambda A(t)]^{2}}|\psi,0\rangle
\end{equation}

\noindent(${\cal T}$ is the time-ordering operator).  This is in the 
``collapse interaction picture" where the operator  
$A(t)\equiv \exp(iH_{A}t)A\exp-(iH_{A}t)$ evolves according to the usual Schr\"odinger 
dynamics and the statevector evolves only due to collapse dynamics. 

	In addition to (\ref{3}), we need to know the 
probability density for each $w(t)$: 

\begin{equation}\label{4}
	{\cal P}_{T}(w)\equiv
	\ _{w}\langle\psi,T|\psi,T\rangle_{w}\, .
\end{equation}

\noindent The probability that $w(t)$ lies between $w(t)$ and $w(t)+dw(t)$ 
for each $t$ in the range $(0,T)$ is 

\begin{equation}\label{5}
	Dw{\cal P}_{T}(w)\equiv
	\prod_{t=0}^{t=T}\frac{dw(t)}{\sqrt{2\pi\lambda/dt}}{\cal P}_{T}(w).
\end{equation}

\noindent (In expressions like (\ref{5}), $t$ may be thought of taking on closely 
spaced discrete values: ${\cal P}_{T}(w)$ is a functional of $w(t)$ for all $0\leq t\leq T$.)  
Since (\ref{3}) is a nonunitary evolution it does not 
preserve statevector norm (which is perfectly all right since, in a collapse theory, 
the direction of a statevector in Hilbert space is all that is needed to 
describe the associated physical reality) so (\ref{4}) says that 
statevectors which have largest norm are most probable.  

	Eqs. (\ref{3}) and (\ref{4}) comprise the CSL model discussed here.  
To see how they work, neglect the unitary evolution (set $H_{A}=0$) and 
set $|\psi,0\rangle=\sum_{i}c_{i}|a_{i}\rangle$ 
($A|a_{i}\rangle=a_{i}|a_{i}\rangle$).  Eqs. (\ref{3}) and (\ref{4}) respectively become 

\begin{eqnarray*}
	|\psi,T\rangle_{w}&
	=&\sum_{i}c_{i}|a_{i}\rangle e^{-\frac{1}{4\lambda}\int_{0}^{T}
 dt[w(t)-2\lambda a_{i}]^{2}}\\
	{\cal P}_{T}(w)&
	=&\sum_{i}|c_{i}|^{2}e^{-\frac{1}{2\lambda}\int_{0}^{T}
 dt[w(t)-2\lambda a_{i}]^{2}}\, .
\end{eqnarray*}

\noindent First, suppose $w(t)=2\lambda a_{j}$.  Then, as $T\rightarrow\infty$,

\begin{eqnarray*}
	|\psi,T\rangle_{w}&\rightarrow &c_{j}|a_{j}\rangle +
	\sum_{i\neq j}c_{i}|a_{i}\rangle e^{-\lambda T [a_{j}-a_{i}]^{2}}\\
	{\cal P}_{T}(w)&\rightarrow &|c_{j}|^{2}+
	\sum_{i\neq j}|c_{i}|^{2} e^{-2\lambda T [a_{j}-a_{i}]^{2}}\, . 
\end{eqnarray*}

\noindent In this case the statevector asymptotically 
"collapses" to $|a_{j}\rangle$. More generally, only if 
$w(t)=2\lambda a_{j}+w_{0}(t)$, where $w_{0}(t)$ is a sample white noise function with 
zero drift (${\cal P}(w_{0})=\exp -(1/2\lambda)\int dtw_{0}(t)^{2}$) will there be a  
non--negligible probability for large $T$.  In that case, 
each of these $|\psi,T\rangle_{w}\rightarrow |a_{j}\rangle$ for large $T$ and the 
total probability of these $w(t)$'s is 
$\int Dw{\cal P}_{T}(2\lambda a_{j}+w_{0}(t))\rightarrow |c_{j}|^{2}$. 
This is, of course, the same result as would be obtained by applying 
the collapse postulate to the original statevector. 

	The density matrix follows from (\ref{3}) and (\ref{4}):  
	
\begin{eqnarray}\label{6}
	\rho (T)&\equiv&\int Dw{\cal P}_{T}(w)\frac{|\psi,T\rangle_{ww}
	\langle\psi,T|}{\ _{w}\langle\psi,T|\psi,T\rangle_{w}}
	=\int Dw|\psi,T\rangle_{ww}\langle\psi,T|\nonumber\\
	&=&{\cal T}e^
	{-\frac{\lambda}{2}\int_{0}^{T} dt[A(t)\otimes 1-1\otimes A(t)]^{2}}\rho (0)
\end{eqnarray} 

\noindent We are employing the notation $(X\otimes Y)Z=XZY$ 
and ${\cal T}$ time--orders operators to the left of $\rho (0)\equiv |\psi,0\rangle 
	\langle\psi,0|$ and 
time--reverse orders operators to the right of $\rho (0)$ so, for example,  Eq. (\ref{3}) in 
a less compact notation is 

\begin{eqnarray*}
	\rho (T) = \rho (0)-\frac{\lambda}{2}{\cal T}\int_{0}^{T} dt[A(t),[A(t), \rho (0)] + ...
\end{eqnarray*}

\noindent The collapse 
behavior in the previous example is easy to see in the 
density matrix (\ref{6}), 

\begin{displaymath}
\rho (t)=\sum_{i}\sum_{j}c_{j}^{*}c_{i}e^{-\frac{\lambda}{2} T(a_{i}-a_{j})^{2}}
|a_{i}\rangle\langle a_{j}|,
\end{displaymath}

\noindent whose off--diagonal elements vanish as $T\rightarrow\infty$. 

 \section{Energy Expectation Conservation}\label{Energy Conservation}
 
\hspace{\parindent}Although $w(t)$ is a classical field, in order to put it on a par with 
 the quantized quantity $A(t)$ we introduce the functional differential 
 operators $\overrightarrow\Pi (t)\equiv i^{-1}\overrightarrow\delta /\delta w(t)$, 
 $\overleftarrow\Pi (t)\equiv -i^{-1}\overleftarrow\delta /\delta w(t)$ (which 
 act on functionals of $w(t)$ to the right 
 and left respectively) and
 $\Pi (t)\equiv\frac{1}{2} (\overrightarrow\Pi (t) + \overleftarrow\Pi (t))$, all of 
 which are conjugate operators to $w(t)$ (i.e., 
 $[w(t),\overleftarrow\Pi (t')]=i\delta (t-t')$, etc.). We define an energy 
 operator for the classical field 
 
  \begin{equation}\label{7}
{\cal H}_{w}\equiv \frac{1}{2}\int_{\infty}^{\infty}dt
[\dot{w}(t)\overrightarrow\Pi (t) + \overleftarrow\Pi (t)\dot{w}(t)]\, .
\end{equation}

	It is readily verified that 
	
\begin{displaymath}
[\frac{d^{n}}{dt^{n}}w(t),{\cal H}_{w}]=i\frac{d^{n+1}}{dt^{n+1}}w(t),\quad 
[\frac{d^{n}}{dt^{n}}\Pi (t),{\cal H}_{w}]=i\frac{d^{n+1}}{dt^{n+1}}\Pi (t) 	
\end{displaymath}

\noindent so ${\cal H}_{w}$ is the time--translation generator for $w(t)$ and $\Pi (t)$.  

	The expectation value of the field energy for the state $|\psi,T\rangle _{w}$ is defined to be 
	
\begin{equation}\label{8}
\bar{\cal H}_{w}(T)\equiv \frac{\ _{w}\langle \psi,T|{\cal H}_{w}|\psi,T\rangle _{w}}
{\ _{w}\langle \psi,T|\psi,T\rangle _{w}}
\end{equation}

\noindent which is real and which 
vanishes at $T=0$ ($|\psi,0\rangle$ does not depend upon $w$,  
so $\overrightarrow\Pi (t)|\psi,0\rangle =0$). It 
should be emphasized that, in spite of the notation, (\ref{8}) is
not a Hilbert space expectation value of ${\cal H}_{w}$: the statevector (\ref{3}) is a vector in $A$'s 
Hilbert space but a functional of $w(t)$.  For example, in (\ref{8}) we cannot 
replace $\overrightarrow\Pi (t)$ by $\overleftarrow\Pi (t)$ as we shall be able to 
do in Eq. (\ref{10}) et. seq. below.  (An analogy is that $\psi^{*}(x)(id\psi(x)/dx)\neq (id\psi(x)/dx)^{*}\psi(x)$ 
but the integrals of both sides {\it are} equal.)

	We also introduce the interaction energy $V\equiv 2\lambda \Pi (0) A$. 
The Schr\"odinger picture's constant energy operator 
$H\equiv H_{0} +V$ ($H_{0}\equiv H_{A}+{\cal H}_{w}$) 
 becomes $H(T)\equiv H_{0}+2\lambda \Pi(T) A(T)$  
($V(T)=\exp (iH_{0}T)V\exp-(iH_{0}T)$) in the interaction picture. 
The expectation value of the total energy for the state $|\psi,T\rangle _{w}$ is

\begin{equation}\label{9}
\bar{H}(w,T)\equiv \frac{\ _{w}\langle \psi,T|H(T)|\psi,T\rangle _{w}}
{\ _{w}\langle \psi,T|\psi,T\rangle _{w}}\, .
\end{equation}

	We now prove the constancy of the expectation value of the ensemble energy 
	
\begin{eqnarray}\label{10}
{\bf \bar{H}}(T)&\equiv&\int 
Dw \ _{w}\langle \psi,T|\psi,T\rangle _{w}\bar{H}(w,T)\nonumber\\
&=&\int Dw \ _{w}\langle \psi,T|H(T)|\psi,T\rangle _{w}\, .
\end{eqnarray}

\noindent Eq. (\ref{10}) is now a scalar product in a Hilbert space in 
which ${\cal H}_{w}$ and $\Pi$ act, and so we may set 
$\overleftarrow\Pi (t)=\overrightarrow\Pi (t)=\frac{1}{2}\Pi (t)$.  Since 
$[\exp-i\int dt B(t)\Pi (t)]f[\int dt w(t)]=f[\int dt (w(t)-B(t))]$ because 
$\Pi (t)$ is the translation operator for $w(t)$,  
we may write the statevector (\ref{3}) as 

\begin{eqnarray}\label{11}
|\psi,T\rangle _{w}&=&{\cal T}e^{-i2\lambda\int_{0}^{T}dtA(t)\Pi (t)}
|\psi,0\rangle e^{-\frac{1}{4\lambda}\int_{0}^{T}dtw^{2}(t)}\nonumber\\
&=&e^{iH_{0}T}e^{-iHT}|\psi,0\rangle 
e^{-\frac{1}{4\lambda}\int_{0}^{T}dtw^{2}(t)}\, .	
\end{eqnarray}

\noindent The last step invokes the well known representation 
of the interaction picture time evolution operator. Since 
$H(T)=\exp(iH_{0}T)H\exp(-iH_{0}T)$, putting (\ref{11}) into (\ref{10}) results in 

\begin{eqnarray}\label{12}
{\bf \bar{H}}(T)&\equiv&\int 
Dw e^{-\frac{1}{4\lambda}\int_{0}^{T}dtw^{2}(t)}\langle \psi,0|
e^{iHT}He^{-iHT}|\psi,0\rangle e^{-\frac{1}{4\lambda}\int_{0}^{T}dtw^{2}(t)}\nonumber\\
&=&\int Dw e^{-\frac{1}{4\lambda}\int_{0}^{T}dtw^{2}(t)}
\langle \psi,0|[H_{A}+{\cal H}_{w}+2\lambda A \Pi(0)]|\psi,0\rangle 
e^{-\frac{1}{4\lambda}\int_{0}^{T}dtw^{2}(t)}\nonumber\\
&=&\langle \psi,0|H_{A}|\psi,0\rangle ={\bf \bar{H}}(0)
\end{eqnarray}

\noindent where the integrals 
$\int Dw\{1,\dot{w}(t)w(t),w(0)\}\exp-(1/2\lambda)\int_{0}^{T}dtw^{2}(t)=\{1,0,0\}$ 
were used in the last step. 

	Thus the energy expectation value is independent of time.  

 \section{Expectation Values of Field and System Energies}\label{Expectation Values}
 
\hspace{\parindent}Here we give expressions for the separate pieces that 
make up ${\bf \bar{H}}(T)$.

	The system energy expectation may immediately be found from (\ref{6}):
	
\begin{equation}\label{13}
{\bf \bar{H}}_{A}(T)={\rm Tr} H_{A}\rho(T)
={\rm Tr}H_{A}{\cal T}e^
	{-\frac{\lambda}{2}\int_{0}^{T}[A(t)\otimes 1-1\otimes A(t)]^{2}}
	|\psi,0\rangle\langle \psi,0|\, .
\end{equation}

	Next we show that ${\bf \bar{V}}(T)=0$. Since 

\begin{eqnarray*}
V(T)|\psi,T\rangle_{w}& =& 2\lambda A(T)\frac{1}{i}\frac{\delta\quad}{\delta w(T)}
{\cal T} e^{-\frac{1}{4\lambda}\int_{0}^{T}dt[w(t)-2\lambda A(t)]^{2}}|\psi,0\rangle\\
& =&iA(T)[w(T)-2\lambda A(T)]|\psi,T\rangle_{w}	
\end{eqnarray*}

\noindent it is straightforward to perform the integral 

\begin{eqnarray*}
{\bf \bar{V}}(T)= \int Dw\ _{w}\langle \psi,T|iA(T)[w(T)-2\lambda A(T)]|\psi,T\rangle_{w}=0
\end{eqnarray*}

\noindent  (because  
$T$ is the largest time, $A(T)$ is at the outside of the time ordering of each statevector, 
so that the average value of $w(T)$ is $2\lambda A(T)$). 

	Last we calculate ${\bf \bar{H}}_{w}(T)$:
	
\begin{eqnarray}\label{14}
{\bf \bar{H}}_{w}(T)
&=&\int Dw\ _{w}\langle \psi,T|\frac{1}{2}\int_{\infty}^{\infty}dt
[\dot{w}(t)\overrightarrow\Pi (t) + \overleftarrow\Pi (t)\dot{w}(t)]|\psi,T\rangle_{w}\nonumber\\
&&\!\!\!\!\!\!\!\!\!\!\!\!\!\!\!\!\!\!\!\!\!\!\!\!\!\!\!\!\!\!
\!\!\!\!\!=\frac{-1}{4\lambda i}\int Dw\int_{0}^{T}dt\dot{w}(t) 
\biggl\{\ _{w}\langle \psi,T|{\cal T}[w(t)-2\lambda A(t)]
e^{-\frac{1}{4\lambda}\int_{0}^{T}dt[w(t)-2\lambda A(t)]^{2}}|\psi,0\rangle\nonumber\\
&&\qquad - \langle \psi,0|{\cal T}_{R}
e^{-\frac{1}{4\lambda}\int_{0}^{T}dt[w(t)-2\lambda A(t)]^{2}}
[w(t)-2\lambda A(t)]|\psi,T\rangle_{w} \biggr\}\nonumber\\
&=&\frac{1}{2i}\int Dw\int_{0}^{T}dt\dot{w}(t) 
\biggl\{\ _{w}\langle \psi,T|{\cal T}A(t)
e^{-\frac{1}{4\lambda}\int_{0}^{T}dt[w(t)-2\lambda A(t)]^{2}}|\psi,0\rangle\nonumber\\
&&\qquad\qquad\qquad - \langle \psi,0|{\cal T}_{R}
e^{-\frac{1}{4\lambda}\int_{0}^{T}dt[w(t)-2\lambda A(t)]^{2}}
A(t)|\psi,T\rangle_{w} \biggr\}
\end{eqnarray} 

\noindent(${\cal T}_{R}$ is the time-reversal operator). 
To perform the functional integral over $w$ in (\ref{14}), i.e., to find the mean value of  $\dot{w}(t)$, 
we first find the mean value of ${w}(t)$ 

\begin{eqnarray*}
\int Dw\, w(t)_{w}\langle \psi,T|\psi,T\rangle_{w}&=&\\
&&\!\!\!\!\!\!\!\!\!\!\!\!\!\!\!\!\!\!\!\!\!\!\!\!\!\!\!\!\!\!\!\!\!\!\!\!{\rm Tr}{\cal T}\lambda 
[A(t)\otimes 1+1\otimes A(t)]e^
	{-\frac{\lambda}{2}\int_{0}^{T}[A(t)\otimes 1-1\otimes A(t)]^{2}}
	|\psi,0\rangle\langle \psi,0|
\end{eqnarray*}

\noindent and so (\ref{14}) becomes

\begin{eqnarray}\label{15}
{\bf \bar{H}}_{w}(T)={\rm Tr}{\cal T}\frac{\lambda}{2i}\int_{0}^{T}dt\,
[\dot{A}(t)\otimes 1+1\otimes \dot{A}(t)][A(t)\otimes 1-1\otimes A(t)]\times\nonumber\\
\qquad\qquad e^{-\frac{\lambda}{2}\int_{0}^{T}dt'[A(t')\otimes 1-1\otimes A(t')]^{2}}
	|\psi,0\rangle\langle \psi,0|\, .
\end{eqnarray}

\noindent The time ordering of $A(t)$ and $\dot{A}(t)$ must be done carefully. We note 
that the integrals in our expressions are of Stratonovich form, e.g.,
 
\begin{eqnarray*}
\int_{t=0}^{T}\dot{w}(t)w(t)&\equiv&
\sum_{t=0}^{T}(\Delta t)^{-1}[w(t+\Delta t)-w(t)][w(t+\Delta t)+w(t)]\\
&=&\sum_{0}^{T}[w^{2}(t+\Delta t)-w^{2}(t)]= w^{2}(T+\Delta t)-w^{2}(0)	
\end{eqnarray*}
	
\noindent rather than, say, Ito integrals (e.g., replacing $[w(t+\Delta t)-w(t)][w(t+\Delta t)+w(t)]$ 
in the above sum by $[w(t+\Delta t)-w(t)][2w(t)]$ which will not give 
a result depending only on the endpoint values of $w$). Therefore{\hfill} 
\break

\begin{eqnarray}\label{16}
\lefteqn{\!\!\!\!\!\!\!\!\!\!\!\!{\cal T}[\dot{A}(t)\otimes 1+1\otimes \dot{A}(t)][A(t)\otimes 1-1\otimes A(t)]
\nonumber}\\
&&\equiv(\Delta t)^{-1}[(A(t+\Delta t)-A(t))
\otimes 1+1\otimes(A(t+\Delta t)-A(t))]\times\nonumber\\ 
&&\qquad\qquad\frac{1}{2}[(A(t+\Delta t)+A(t))\otimes 1-
1\otimes (A(t+\Delta t)+A(t))]\nonumber\\
&&=(2\Delta t)^{-1}\Bigl\{ (A^{2}(t+\Delta t)-A^{2}(t))\otimes 1
-1\otimes(A^{2}(t+\Delta t)-A^{2}(t))\nonumber\\
&&\qquad\qquad+2[A(t)\otimes A(t+\Delta t)-A(t+\Delta t)\otimes A(t)]\Bigr\}\, .
\end{eqnarray}

\noindent We next note, since  the $A$--terms in (\ref{16}) have upper 
time $t+\Delta t$, that the time-ordering and trace operations in (\ref{15}) yield  

\begin{eqnarray*}
{\rm Tr}{\cal T}e^{-\frac{\lambda}{2}\int_{t+\Delta t}^{T}dt'
[A(t')\otimes 1-1\otimes A(t')]^{2}}=1
\end{eqnarray*}

\noindent  so $t$ may be set as the 
upper limit of the exponential's integral in (\ref{15}).   
Then the $A$--terms in (16) have the largest times in (the thus-modified) (\ref{15}), so that 
the time-ordering and trace operations allow one to apply $B\otimes C=CB$ to 
(\ref{16}). The $A^{2}$ terms in (\ref{16}) cancel, leaving 

\begin{eqnarray}\label{17}
\lefteqn{{\cal T}[\dot{A}(t)\otimes 1+1\otimes \dot{A}(t)][A(t)\otimes 1-1\otimes A(t)]}\nonumber\\ 
&&\qquad=(\Delta t)^{-1}[A(t+\Delta t),A(t)]= [\dot{A}(t),A(t)]=i[A(t),[A(t),H_{A}]]\, .
\end{eqnarray}

\noindent Putting (\ref{17}) into (\ref{15}) gives the desired result:

\begin{eqnarray}\label{18}
{\bf \bar{H}}_{w}(T)\!=\!{\rm Tr}{\cal T}\frac{\lambda}{2}\!\int_{0}^{T}\!\!\!dt\,
[A(t),[A(t),H_{A}]]
e^{-\frac{\lambda}{2}\int_{0}^{t}dt'[A(t')\otimes 1-1\otimes A(t')]^{2}}
	|\psi,0\rangle\langle \psi,0|\, .
\end{eqnarray}

\section{Concluding Remarks}\label{Conclusion}

\hspace{\parindent}It can immediately be seen from (\ref{13}) and (\ref{18}) that 
	
\begin{eqnarray}\label{19}
&&\!\!\!\!\!\!\!\!\!\!\!\!\!\!\!\!\!\!\!\!\!\!\!\!\!\!\!
\frac{d}{dT}{\bf \bar{H}}_{A}(T)=-\frac{d}{dT}{\bf \bar{H}}_{w}(T)\nonumber\\
&&=-\frac{\lambda}{2}{\rm Tr}{\cal T}[A(T),[A(T),H_{A}]]
e^{-\frac{\lambda}{2}\int_{0}^{T}dt'[A(t')\otimes 1-1\otimes A(t')]^{2}}
	|\psi,0\rangle\langle \psi,0|\, .
\end{eqnarray}

	For a simple example, consider a single free particle 
moving in one dimension where collapse is toward a position eigenstate\cite{Diosi 89}, i.e., 
$A=x$ and $A(t)=x+(p/m)t$. Since $[x(t)[x(t),p^{2}/2m]]=-\hbar^{2}/2m$, 
it follows from (\ref{19}) that 

\begin{eqnarray*}
\frac{d}{dT}{\bf \bar{H}}_{A}(T)=-\frac{d}{dT}{\bf \bar{H}}_{w}(T)=
\frac{\lambda \hbar^{2}}{2m}
\end{eqnarray*}

\noindent so, from the initial values of (\ref{13}), (\ref{18}),

\begin{eqnarray*}
{\bf \bar{H}}_{A}(T)&=&\langle\psi,0|H_{A}|\psi,0\rangle + \frac{\lambda \hbar^{2}}{2m}T\\
{\bf \bar{H}}_{w}(T)&=&-\frac{\lambda \hbar^{2}}{2m}T\, . 
\end{eqnarray*}

\noindent Thus, since collapse narrows wavefunctions (toward 
eigenstates of position), the ensemble average particle energy steadily increases 
at the expense of a steadily decreasing ensemble average field energy. 

	It is straightforward to apply the results given here to nonrelativistic CSL, 
by replacing $(t)$ by $({\bf x},t)$, e.g., the expressions now contain 
$w({\bf x},t)$, $\Pi({\bf x},t)$, 
$A({\bf x},t)$ (the particle number density operator smeared by a gaussian of width 
$\approx 10^{-5}$ cm), $d{\bf x}dt$, etc. One may also readily obtain conservation of the 
expectation values of the other geometric conservation laws such as momentum

\begin{displaymath}
{\cal P}_{w}\equiv \frac{1}{2}\int_{-\infty}^{\infty}d{\bf x}dt
[{\bf \nabla}w({\bf x},t)\overrightarrow{\Pi}({\bf x},t)+
\overleftarrow{\Pi}({\bf x},t){\bf \nabla}w({\bf x},t)]\,.
\end{displaymath} 

	To conclude, we reiterate that this calculation gives a bit of support to the 
suggestion that conservation law violation obtained in applying SQT's collapse 
postulate could be overcome in a dynamical collapse theory.  

\eject

\end{document}